\documentclass[11pt,twoside]{article}
\usepackage{asp2004}
\usepackage{psfig}
\usepackage{epsf}
\usepackage{graphics}
\usepackage{lscape}
\def\edcomment#1{\iffalse\marginpar{\raggedright\sl#1\/}\else\relax\fi}
\reversemarginpar

\begin{document}
\title{Star Formation in the W49A Molecular Cloud: Birth of a Massive Star Cluster}
\author{N. L. Homeier \& J. Alves}
\affil{(1) Johns Hopkins University, Baltimore, MD, USA\\ (2) European Southern Observatory, Garching, Germany}

\begin{abstract}
The W49A star-forming region is embedded in a $10^{6} M_{\odot}$ molecular 
cloud, one of the most massive in our Galaxy. It has been long known as one 
of the most luminous radio H~II regions, containing $30-40$ compact and 
ultracompact H~II regions and several hot cores. We 
have detected a previously unknown massive star cluster (Cluster~1) embedded 
in the W49 molecular cloud using $JHKs$ observations with SOFI+NTT. 
We find that the inferred mass of Cluster 1 is $1-2 \times 10^4$~M$_{\odot}$, 
and is 2~pc in 
projected distance from the largest grouping of ultracompact H~II regions 
(including the Welch ring). 
We use the extensive line-of-sight extinction to isolate a population
of objects associated with W49A, and use this sample to 
obtain a mass function. The slope of the 
derived mass function for objects associated with W49A, $-1.3 \pm 0.3$, is 
consistent with a Salpeter slope. 
About 3~pc away from the main star-forming complexes
seen in near-infrared and radio observations is an $\sim 80$~M$_{\odot}$ 
star ionizing a compact H~II region (object CC). We obtained
adaptive optics imaging with NACO on the VLT of the 1.5~pc surrounding 
this object to search investigate the stellar initial mass function 
in the vicinity of a massive star.
On the global molecular cloud scale in W49, massive star formation apparently 
did not proceed in a single
concentrated burst, but in small groups, or subclusters. 

\end{abstract}
\thispagestyle{plain}

\section{Introduction}

The W49 Giant Molecular Cloud extends more than 50~pc
in diameter and weighs in at $\sim 10^6$
M$_\odot$ \cite{Setal01}, the most massive in our Galaxy outside of 
the Galactic
Center. Embedded within this cloud, W49A 
is one of the most luminous Galactic giant radio H~II regions
($\sim 10^7 L_\odot$). The W49A star-forming
region lies in the Galactic plane 
%($l = 43.17^\circ, b =+0.00^\circ$) 
at a distance of 11.4$\pm$1.2 kpc \cite{Getal92}, has
$\sim$40 well studied UC~HII regions (e.g. DePree et~al. 1997), 
associated with a minimum of that number
of central stars earlier than B3 \cite{C02}. About 12 of these radio
sources are arranged in the well known Welch ``ring'' \cite{welch87}.
A few other young Galactic clusters have a large number of massive
stars, e.g., the Carina nebula (e.g. Rathborne et~al. 2002), 
NGC~3603 (e.g. Drissen et~al. 1995), and the Arches cluster
(e.g. Figer et~al. 2002), or
are very young, e.g. NGC~3567 \cite{Betal03,Fetal02}, W42
and W31 (e.g. Blum et al. 2001) but no other known
region has a large number of massive stars in such a highly embedded
and early evolutionary state.  For this reason W49A is unique in our
known Galaxy.  

\section{Star Formation in the W49A Molecular Cloud}

Our images contain many stars along the line of sight, but we can
use the reddening within the Galactic disk to our advantage.
We identify a stellar population associated with the W49A 
region by first selecting objects with $H-Ks$
colors red enough to be consistent with a distance of $11.4$~kpc along the
Galactic plane, estimated by assuming an
exponential distribution of Galactic dust. We arrive at A$_{K}=2.1$ and 
$H-K=1.2$ (Rieke~\&~Lebofsky~(1985) 
reddening law).

To get an unbiased luminosity function for the stars associated with 
W49, so we select an extinction-limited sample of stars, with 
limits set by the extinction and completeness limits (see Homeier \& 
Alves 2004)
As the sample suffers from severe non-uniform extinction, we 
corrected for this effect assuming an intrinsic color of $H-Ks=0.15$. 
This was chosen based on the knowledge 
that all stars without hot dust are intrinsically 
nearly colorless in the near-infrared, with $H-Ks$ ranging from 0.0 to 0.3. 
We expect objects associated with the W49A star-forming region to be early-type
stars with intrinsic $H-Ks$ near 0.0, whereas giant stars should
have intrinsic $H-Ks$ up to 0.3. 

\begin{figure}
\plotfiddle{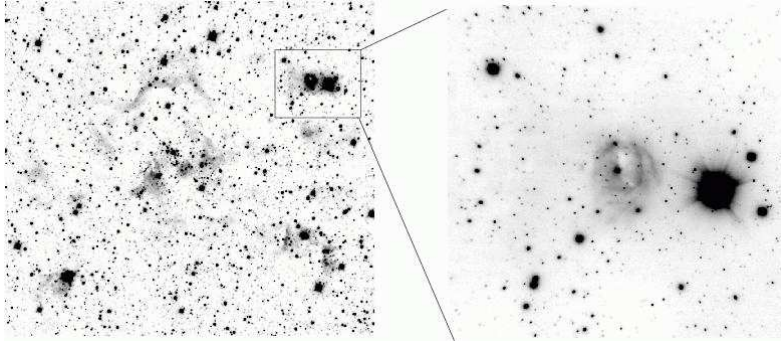}{1.6in}{0}{42}{42}{-170}{-7}
\caption{A SOFI $Ks$ image of the W49A region is shown in the left panel, 
with the NACO $Ks$ image of the 'CC' object in the right panel.
The SOFI image is $5\arcmin\times4\arcmin$ ($\sim16\times16$~pc),
and the NACO image is $50\arcsec\times50\arcsec$.
\label{fig1}}
\end{figure}

We convert the extinction-limited extinction-corrected 
$Ks$ luminosity function for our entire field into a mass 
function by transforming each magnitude bin to a mass. 
The relationship between mass and absolute
magnitude is taken from the $4\times10^{5}$~yr isochrones of 
Lejeune~\&~Schaerer (2001).
We extrapolated the magnitude-mass relation to infer masses for the most
luminous stars, which are more luminous than the 120~M$_{\odot}$ models.
The error in the slope is large and there are many sources of 
uncertainty in our measurement, but we can say that we do not see 
evidence for a top-heavy IMF. We find a mass function 
slope of $-1.3\pm0.3$, which is
consistent with Salpeter.

\subsection{Total Stellar Mass}

We estimate the total stellar mass of the W49A star
cluster by counting stars with masses greater than $20$~M$_{\odot}$ 
and using a Salpeter slope, with upper and
lower mass limits as 120~M$_{\odot}$ and 1~M$_{\odot}$. 
For Cluster~1, we
find 54 stars within $45 \arcsec$, implying a total mass of 
$\sim 1 \times 10^{4}$~M$_{\odot}$. In our entire field, we 
count 269 stars with masses $\ge 20$~M$_{\odot}$, implying a
total mass of $5-7 \times 10^{4}$~M$_{\odot}$. The stars we
have identified as massive stars are
certainly contaminated by background obejcts,
but we are also certainly incomplete in our census of massive
stars due to extinction and angular resolution. Even if the stellar 
mass estimate for W49A is a factor of 2 too high, W49A is as or 
more massive than any known young Galactic star cluster. 
It is important to note that this is a {\it lower limit} 
to the final stellar mass,
as there is circumstantial and direct evidence for ongoing star 
formation in this region. There is abundant molecular gas,
and hot cores near the ring of UC~HII regions \cite{Wetal01}. 

\subsection{NACO observations: Low-mass star formation near a $60-80$ M$_{\odot}$ star and a popped bubble}

We obtained $JHKsL$ adaptive optics imaging observations with NACO 
on the  VLT of the
$1\arcmin\times1\arcmin$ (3pc$\times$3pc) region surrounding the
'CC' object \cite{depree97}, which is appears to be a single 
(to 600~AU) massive star, $60-80$ M$_{\odot}$, ionizing a compact 
HII region. Our objective is to search for the occurence of 
low-mass star formation in the vicinity of this high mass star. 
To accomplish this, we look for evidence for clustering of 
sources with extinction consistent with association with the 
W49A star-forming region. 
We find tentative evidence for an excess of sources within
$0.5-1.0$~pc of the CC star, however, this is a work in progress
and needs to be mroe carefully examined for the statistical
significance.

The radius of the compact HII region surrounding the CC star
is $0.1-0.2$~pc. Our NACO NIR images show that the HII region 
has broken out of the confining material on the eastern side. 
We detect diffuse emission extending away from the star and HII 
region, which is most likely due to strong nebular lines of H in 
the $H$ and $Ks$ passbands. This is supported by the 3.6cm radio 
observations by de~Pree et~al. (1997), in which the ionized 
gas shows the same morphology. This ionized emission extends 
over 10 times the radius of the HII region, to $\sim1$~pc.

\section{Conclusions}

Some of the NIR sources associated with W49A were 
detected by Conti \& Blum
(2002), but they did not have a large enough FOV to identify
a NIR cluster with the radio HII region.
Our observations clearly show a massive star cluster 
adjacent to the UC~HII regions (2~pc distant)
\cite{AH03}.
This means that the W49A region began forming stars 
earlier than previously thought, and that the
ultra-compact HII\ regions which have been long-known to radio astronomers
are not the first generation of massive stars. 
We use these data to estimate a total stellar mass of 
$5-7\times10^{4}$~M$_{\odot}$. Since molecular gas is
abundant, this is a lower limit to the final stellar mass of the cluster.
With these observations, W49A joins the list of Galactic giant
radio HII regions with the coexistence of two or more phases of 
massive star formation. This means that the formation of 
massive stars in clusters is not completely synchronized, but 
there is some spread in age. 

What does the core of Cluster~1 hold? Given the high internal extinction, 
we are certainly incomplete in our near-infrared 
census of star formation and therefore any mass or density estimates. 
The core is
very crowded; high spatial resolution observations are needed. Taken at 
face value and without correcting for the large extinction,
the cluster core appears to be 
significantly less dense 
than the Arches, NGC~3603, or 30~Doradus. If it is truly less dense,
then the different formation environment of  
W49A may be an important clue for understanding the processes
which drive clustered star formation.

The subclustering phenomenon is useful in describing the 
star formation pattern in the W49A molecular cloud. 
When the cloud has ceased forming stars, the resulting 
stellar group will likely be called a 'cluster', but at the
time of current observation, the massive star formation 
does not appear to be distributed uniformly throughout the region,
or with a radial dependence relative to a cluster 'center'. 
Rather it is better described as occuring in 'subclusters'. 
In this sense we could count $4-5$ subclusters within $\sim 13$~pc
using the combined NIR and radio
observations: Cluster~1, the (Welch) ``ring'' of UC~HII
regions, W49A South, the RQ complex, and perhaps the CC source.
We speculate that star formation within a subcluster is essentially 
synchronized, and a massive star cluster is a collective of
several (or many) subclusters. There is also evidence for
subclustering in lower-mass star-forming regions \cite{LAL96,Tetal00},
which can be reproduced in star formation simulations \cite{BBV03}.
Possible examples of subclustering in
extragalactic star clusters are: SSC-A in NGC~1569 and NGC 604 in M33. 
SSC-A in NGC~1569 has a
stellar concentration with red super giants and another with 
Wolf-Rayet stars.
The massive stars in NGC 604 are certainly subclustered, but the
region itself it is of sufficiently 
low density to be termed a Scaled OB Association (SOBA) rather
than a star cluster \cite{M-A01}. The applicability of the subclustering 
description to other young massive Galactic star clusters remains to be seen,
but we conclude that it is a useful concept to describe and understand massive
star formation in W49A.

\end{document}